# Comment on "New proof of general relativity through the correct physical interpretation of the Mössbauer rotor experiment" by C. Corda

Alexander L. Kholmetskii [1], Tolga Yarman[2], Ozan Yarman[3] and Metin Arik[4]

[1]Department of Physics, Belarus State University, Minsk, Belarus, tel. +375 17 2095482, fax +375 17 2095445, e-mail: alkholmetskii@gmail.com
[2]Okan University, Istanbul, Turkey & Savronik, Eskisehir, Turkey
[3]Istanbul University, Istanbul, Turkey
[4]Bogazici University, Istanbul, Turkey

**Abstract.** We analyze the attempt by C. Corda to explain the results of modern Mössbauer experiments in a rotating system via the additional effect of synchronization of the clock in the origin of the rotating system with the laboratory clock, and indicate errors committed by him.

In a recent paper by C. Corda [1] – which essentially repeats his previous publication [2] – the author once more claims that the outcomes of recent experiments conducted by our team towards the measurement of the Mössbauer effect in a rotating system [3-6] represent "…*a new, strong and independent, proof of general relativity*".

We remind that in these experiments, where a source of resonant radiation and a resonant absorber are both fixed on a rotor, the linear Doppler effect between the source and the absorber does not emerge; whereas, the relative second order Doppler shift, written to the accuracy of calculations up to $c^{-2}$, is given by the general expression

$$\frac{\Delta E}{E} = -k\frac{u^2}{c^2}, \qquad (1)$$

where the sign "minus" corresponds to the configuration, when a source of resonant radiation is located on the rotor axis, while a resonant absorber is fixed on the rotor rim, with $u$ being its tangential velocity (see Fig. 1). It was believed for a long time that the coefficient $k$ is solely defined by the relativistic time dilation effect for the orbiting absorber, which yields $k$=0.5 in eq. (1) (see, e.g. [7]).

This result was presumed to have been confirmed in Mössbauer rotor experiments performed during the mid-20[th] century. However, as we have shown in [8], the rectified reprocessing of the data published by Kündig [9] and Champeney et al. [10], motivated by the original prediction made by T. Yarman [11], rather leads to the inequality

$$k \geq 0.6, \qquad (2)$$

and the difference from the relativistic prediction $k$=0.5 exceeds by at least 10 times the measurement uncertainty reported by the authors of these experiments (≤0.01).

The finding (2) had been successfully verified in two recent experiments conducted by our team, thence leading to

$$k=0.66\pm0.03 \text{ [3, 4]} \qquad (3)$$

and

$$k=0.69\pm0.02 \text{ [5, 6]}. \qquad (4)$$

The results (2)-(4) indicate that the measured energy shift between an emitted and a received radiation in a rotating system is defined not only by the ordinary relativistic dilation of time for an orbiting resonant absorber, but does furthermore entail an additional effect, which is responsible for the extra-energy shift (hereinafter abbreviated as the EES) between emission and absorption lines; thus constituting about 30 % deviation from the expected relativistic value $k$=0.5.

These findings stimulated scientists for a search of the physical origin of the EES, and one of the first attempts to provide its physical interpretation had been presented by C. Corda in ref. [2], and now rehashed in the present publication [1]. According to him, all preceding analyses of Mössbauer rotor experiments, which predicted the value $k$=0.5 in eq. (1), missed the effect of a clock synchronization between the spinning source (mounted on the rotational axis) and a detector of γ-quanta (located outside the rotor system). Thus, the detector is moving with respect to the origin of a rotating frame; therefore, according to Corda, for a correct determination of the coefficient $k$ in eq. (1), the clock in the detector must have been synchronized with the clock in the origin; which should give (according to him) an additional contribution to the energy shift between the source and the detector. Based on his calculations in [1, 2], this additional component of the relative energy shift is given by the equation

$(\Delta E/E)_{\text{source-detector}} = -u^2/6c^2$, (5)

which therefore should (according to Corda) be added to the conventional relative energy shift between the lines of the resonant source and the absorber due to the time dilation effect

$(\Delta E/E)_{\text{source-absorber}} = -u^2/2c^2$. (6)

Thereby, in Corda's mindset, the total relative energy shift measured in the Mössbauer rotor experiment is to be defined as the sum of the energy shift components (5) and (6); i.e.,

$(\Delta E/E)_{\text{total}} = (\Delta E/E)_{\text{source-absorber}} + (\Delta E/E)_{\text{source-detector}} = -u^2/2c^2 - u^2/6c^2 = -2u^2/3c^2$. (7)

A comparison of eqs. (1) and (7) yields $k=2/3$, which seemingly agrees with the results of the experiments (3), (4).

Based on eq. (7), Corda concluded that the results (3), (4) represent a "*new, strong and independent proof of the correctness of Einstein's vision of gravity*" [2].

However, we stress that the summation of the energy shift components (5) and (6) would be legitimate only in the case of an equal sensitivity of the detection system to both kinds of energy shifts, which obviously is not the case for Mössbauer rotor experiments [12]. Further discussion on this subject [13, 14] – hidden by Corda in his paper [1] – well confirmed our conclusion that the energy shift component (6) cannot be measured in this kind of experiments.

By this time, we see no further need to reproduce our results contained in refs. [12, 14]. In what follows, we finalize the entire discussion on the subject and show that the "synchronization effect" by Corda results from evident mathematical errors, he regrettably committed.

In order to indicate said errors, we would like to foremost of all remind that in the calculation of the so-called synchronization effect between the clock of detector and the clock in the origin of a rotational system, C. Corda referred to the paper by Ashby [15], where the transformation between an inertial frame and a rotating frame in cylindrical coordinates is given by the relationships

$t = t'$, $r = r'$, $\phi = \phi' + \omega t'$, $z = z'$, (8a-d)

with $\omega$ being the angular rotation frequency. Hereinafter, the non-primed quantities are referred to the inertial (laboratory) frame, while the primed quantities are referred to the rotating frame. Thus, using the expression for the space-time interval in cylindrical coordinates of an inertial reference frame

$ds^2 = c^2 dt^2 - dr^2 - r^2 d\phi^2 - dz^2$, (9)

and combining eqs. (8), (9), one obtains the following expression for the space-time interval in the rotating frame (Langevin metric) [15]

$ds^2 = \left(1 - \dfrac{r'^2 \omega^2}{c^2}\right) c^2 dt'^2 - 2\omega r'^2 d\phi' dt' - dr'^2 - r'^2 d\phi'^2 - dz'^2$. (10)

Hence, one immediately derives the expression for the proper time increment [15]

$d\tau^2 = dt'^2 \left[1 - \left(\dfrac{\omega r'}{c}\right)^2 - \dfrac{2\omega r'^2 d\phi'}{c^2 dt'} - \left(\dfrac{d\sigma'}{c dt'}\right)^2\right]$, (11)

where the designation $d\sigma'^2 = dr'^2 + (r' d\phi')^2 + dz'^2$ has been used.

Eq. (11) can be straightforwardly applied to the derivation of the time dilation effect between a source of resonant radiation, located on the rotor axis and a resonant absorber, located at the rotor rim with the radius $r'$. Taking into account that, for the resonant absorber, fixed on the rotor, $dr'=d\phi'=dz'=0$, we immediately derive from eq. (11) the time dilation effect for this absorber as

$$d\tau = dt'\sqrt{1-(\omega r'/c)^2}\ .$$

Designating $\omega r' = u$, we obtain within the sufficient accuracy of calculations of the order $(u/c)^2$

$$d\tau \approx dt'(1 - u^2/2c^2), \tag{12}$$

which straightly leads to the first contribution (6) of the relative energy shift between the resonant lines of the source and the absorber.

In order to calculate the second contribution (5) to the relative energy shift – which, according to Corda, stands for his "synchronization effect" that was somehow missed all this time in Mössbauer rotor experiments [1] – he again refers to eq. (11), but unfortunately commits errors.

For the purpose to accentuating these errors, we remind that in the calculation of the synchronization effect between the clock of detector and the clock in the origin of a rotational system, C. Corda has used eq. (10) of ref. [15] that he modified to the form

$$d\tau = dt'\left(1 - \frac{r'^2\omega^2}{c^2}\right) \tag{13}$$

(where $dt'$ is the time increment at $r'$), which, according to Corda [1, 2], "… *represents the proper time increment $d\tau$ on the moving clock having radial coordinate r' for values v<<c.*"

To clarify the meaning of the strong inequality $v<<c$, mentioned by Corda, we referred to the original work by Ashby [15], and found out that the precursor of the equation (13) (eq. (10) of [15]) is derived from the Langevin metric for a very specific case, where extremely slowly (i.e., quasistatically) moving portable clocks are used to disseminate time. In such a case, the parameter $v$ stands for the *velocity* of said portable clock, which should be assumed to be negligibly small; i.e., $v\to 0$. Then, the expression for the proper time increment in the Langevin metric (11) can indeed be substantially simplified. Namely, in the adopted limit $v\to 0$ for a portable clock, Ashby used the *linear* approximation to the ratio ($\omega r'/c$), where he neglected the terms $(d\sigma'/cdt')^2$ and $(\omega r'/c)^2$ in eq. (11), and arrived at

$$d\tau \approx dt' - \frac{\omega r'^2 d\phi'}{c^2}\ . \tag{14}$$

(see eq. (10) of [15]). However, the linear approximation to the ratio ($\omega r'/c$) is obviously inapplicable to Mössbauer rotor experiments, where the measured energy shift between an emitted and an absorbed radiation is proportional to ($\omega r'/c)^2$, and whereby the neglection of the terms $(d\sigma'/cdt')^2$ and $(\omega r'/c)^2$ in eq. (11) is inadmissible. This already indicates that the approach by Corda, which uses eq. (14), is incorrect.

Further, Corda takes into account that the detector of $\gamma$-quanta is moving with respect to the origin of a rotating frame, and adopts that

$$d\phi' = \omega dt'. \tag{15}$$

Hence, substituting eq. (15) into eq. (14), one arrives at eq. (13), which had been used by C. Corda in the calculation of his so-called synchronization effect [1].

However, in addition to the inapplicability of eq. (14) to Mössbauer rotor experiments, as mentioned above, we further highlight that eq. (15) is itself erroneous.

Indeed, the motional equation of the detector in the rotating frame can be found via eqs. (8a-d), where we adopt the constant values of $r$, $\phi$ and $z$ in the laboratory frame, wherein the detector is at rest. Therefore, we get $d\phi=0$ in eq. (8c), and consequently

$$d\phi' = -\omega dt', \tag{16}$$

which thus differs in sign from eq. (15). Thus, instead of eq. (13), we should obtain

$$d\tau = dt'\left(1 + \frac{r'^2\omega^2}{c^2}\right), \tag{17}$$

which obviously changes the numerical estimation of the "synchronization effect" by Corda.

At the same time, there is no meaning to specify the new numerical value of such a "synchronization effect"; it is sufficient to notice that eq. (14), as applied in ref. [15] for a portable clock in the linear approximation to the ratio ($\omega r'/c$), is anyway inapplicable to Mössbauer rotor experiments, where the exact expression for the proper time interval (11) should be used. Additionally, we should take into account that in the rotating frame, the motion of the detector of $\gamma$-radiation is described by the equation (16) under the constant values of $r'$ and $z'$, where

$$dr' = 0, \quad dz' = 0. \tag{18a-b}$$

Thus, substituting eqs. (16) and (18a-b) into the general equation (11), we obtain

$$d\tau^2 = dt'^2 \left[ 1 - \left(\frac{\omega r'}{c}\right)^2 + \frac{2\omega^2 r'^2}{c^2} - \frac{r'^2 \omega^2}{c^2} \right],$$

which yields

$$d\tau = dt'. \tag{19}$$

Consequently, the clock in the origin of a rotating system and a laboratory clock – being synchronized to each other before the rotor run – both stay synchronized at any given angular frequency $\omega$, and the entire "synchronization effect" by Corda completely disappears.

Thus, the problem of the interpretation of the EES still remains open to discussion.

Finally, we point out a false citation by Corda of our papers in ref. [1]. In the conclusion section he writes: "*The importance of the results of this Essay is stressed by the issue that various papers in the literature… missed the effect of clock synchronization [1-8], [11-13], with some subsequent claim of invalidity of relativity theory and/or some attempts to explain the experimental results through "exotic" effects [1, 2, 11, 12, 13], which, instead, must be rejected*".

However, our abovementioned paper referred to as 1 (now ref. [8]) is devoted solely to the re-analysis of the experiments [9] and [10], where the inequality (2) had been found, while our papers referred to as 2, 13 (now refs. [3], [5], correspondingly) describe our Mössbauer rotor experiments and present their experimental results (eqs. (3) and (4)). We emphasize that these papers contain no attempts to explain eqs. (3), (4) via any "exotic effects" or to claim the invalidity of relativity theory, so that the present phrase by Corda only serves to mislead the readers.

Finally, the absence of any "synchronization effect" in the available approaches to explain the origin of the EES (refs. 11, 12 in [1]) represents an advantage rather than a shortcoming.

**Acknowledgment**

We thank our anonymous reviewer for his helpful remarks and suggestions, which allowed us to considerably improve the presentation of this comment.

**References**

[1] C. Corda, Int. J. Mod. Phys. D, DOI: 10.1142/S0218271818470168.
[2] C. Corda, Ann. Phys. **355**, 360 (2015).
[3] A.L. Kholmetskii, T. Yarman, O.V. Missevitch and B.I. Rogozev, Phys. Scr. **79**, 065007 (2009).
[4] A.L. Kholmetskii, T. Yarman and O.V. Missevitch, Int. J. Phys. Sci. **6**, 84 (2011).
[5] A.L. Kholmetskii, T. Yarman, M. Arik and O.V. Missevitch, AIP Conf. Proc. **1648**, 510011 (2015).
[6] T. Yarman, A.L. Kholmetskii, M. Arik, B. Akkus, Y. Öktem, L.A. Susam and O.V. Missevitch, Can. J. Phys. **94**, 780 (2016).
[7] L.D. Landau and E.M. Lifshitz, The Classical Theory of Fields, (Butterworth & Heinemann, 1999).
[8] A.L. Kholmetskii, T. Yarman and O.V. Missevitch, Phys. Scr. **78**, 035302 (2008).
[9] W. Kündig. Phys. Rev. **129**, 2371 (1963).
[10] D.C. Champeney, G.R. Isaak and A.M. Khan, Proc. Phys. Soc. **85**, 583 (1965).
[11] T. Yarman, Ann. Fond. de Broglie **29**, 459 (2004).
[12] A.L. Kholmetskii, T. Yarman and M. Arik, Ann. Phys. **363**, 556 (2015).

[13] C. Corda, Ann. Phys. **368**, 258 (2016).
[14] A.L. Kholmetskii, T. Yarman, O. Yarman, M. Arik, Ann. Phys. **374**, 247 (2016).
[15] N. Ashby, Living Rev. Relativity **6**, 1 (2003).

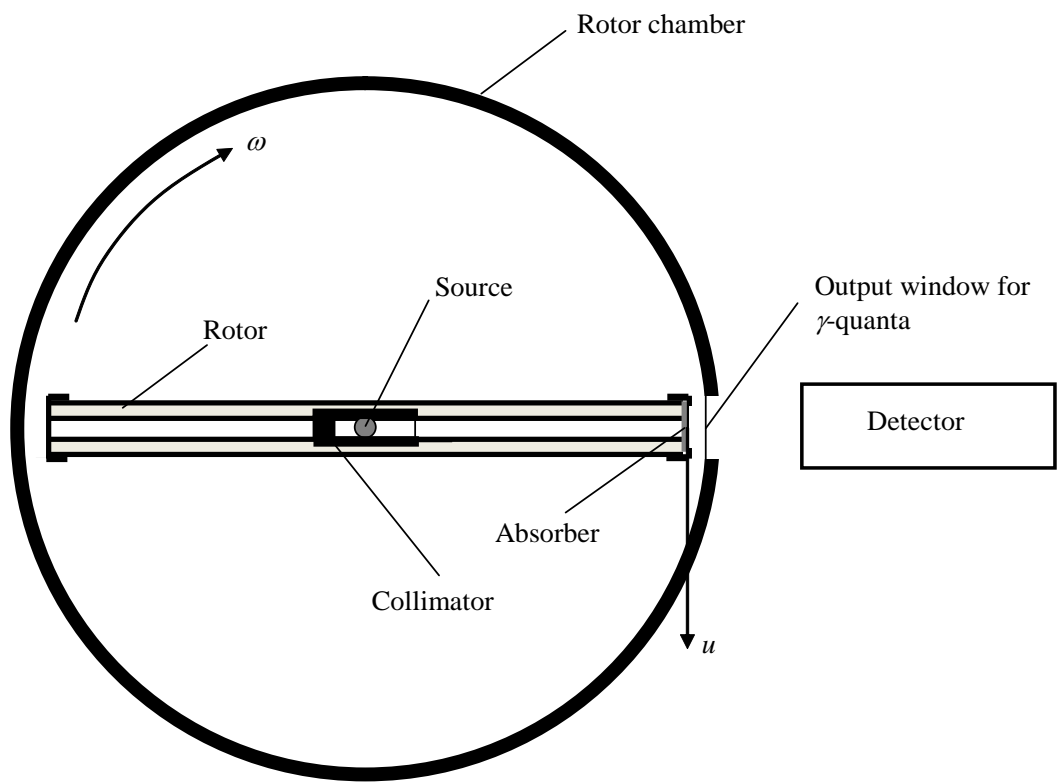

Figure. 1, adapted from ref. [6]. General scheme of the Mössbauer rotor experiment. A source of resonant radiation is fixed on the rotor axis; an absorber is mounted on the rotor rim, while a detector of $\gamma$-quanta is placed outside the rotor system. $\gamma$-quanta emitted by the source and passing across the absorber are detected at the time moments, when source, absorber and detector are aligned in a straight line.